\documentclass[conference]{IEEEtran}
\IEEEoverridecommandlockouts
\usepackage{cite}
\usepackage{amsmath,amssymb,amsfonts}
\usepackage{algorithmic}
\usepackage{graphicx}
\usepackage{textcomp}
\usepackage{xcolor}
\usepackage{bm}
\usepackage{upgreek}
\usepackage{color}
\usepackage{tikz}
\usepackage{pgfplots}
\usepackage{scalerel}
\usepackage{multirow}
\usepackage{makecell}
\usepackage{booktabs}
\usepackage[caption=false]{subfig}
\pgfplotsset{compat=1.7}
\def\BibTeX{{\rm B\kern-.05em{\sc i\kern-.025em b}\kern-.08em
    T\kern-.1667em\lower.7ex\hbox{E}\kern-.125emX}}
\begin{document}

\title{Online Domain-Incremental Learning Approach to Classify Acoustic Scenes in All Locations\\
\thanks{This work was supported by Jane and Aatos Erkko Foundation under grant number 230048, "Continual learning of sounds with deep neural networks". \\
The authors wish to thank CSC-IT Centre of Science Ltd., Finland,  for providing computational resources.} 
}

\author{\IEEEauthorblockN{Manjunath Mulimani}
\IEEEauthorblockA{\textit{Signal Processing Research Center} \\
\textit{Tampere University}\\
Tampere, Finland\\
manjunath.mulimani@tuni.fi}
\and
\IEEEauthorblockN{Annamaria Mesaros}
\IEEEauthorblockA{\textit{Signal Processing Research Center} \\
\textit{Tampere University}\\
Tampere, Finland \\
annamaria.mesaros@tuni.fi}
}

\maketitle

\begin{abstract}
In this paper, we propose a method for online domain-incremental learning of acoustic scene classification from a sequence of different locations. Simply training a deep learning model on a sequence of different locations leads to forgetting of previously learned knowledge. In this work, we only correct the statistics of the Batch Normalization layers of a model using a few samples to learn the acoustic scenes from a new location without any excessive training. Experiments are performed on acoustic scenes from 11 different locations, with an initial task containing acoustic scenes from 6 locations and the remaining 5 incremental tasks each representing the acoustic scenes from a different location. The proposed approach outperforms fine-tuning based methods and achieves an average accuracy of 48.8\% after learning the last task in sequence without forgetting acoustic scenes from the previously learned locations.

\end{abstract}

\begin{IEEEkeywords}
Domain-incremental learning, online learning, acoustic scene classification, Batch Normalization layers, forgetting, deep learning model
\end{IEEEkeywords}

\section{Introduction}
The humans' auditory system can detect the surrounding environments--acoustic scenes, through the analysis of sounds, irrespective of the geographical locations. This shows that humans have the natural ability of lifelong learning and barely forget previously learned audio patterns when moving to new locations or domains worldwide. However, despite the recent success of deep learning models on various machine listening tasks, incremental or continual learning of audio tasks with evolving domains
leads to what is called catastrophic forgetting \cite{parisi2019continual}. It means, 
the performance of the deep learning model on previously learned domains drastically degrades when trained on a new domain, given that the dataset of the previous domains is not accessible.   The change of input distributions in the domains, known as domain or distribution shift,  is the main reason for performance degradation. A slight distribution shift can degrade the performance of the deep learning models significantly \cite{hendrycks2018benchmarking, recht2019imagenet}.

A straightforward approach to avoid forgetting is training a separate model for each static location. However, this requires time-consuming data collection, annotation, and re-training. In addition, it is required to save all the models, and it may not be possible due to computational constraints. 
Unlike conventional deep learning models,  domain-incremental learning (DIL) allows a single model to learn continuously evolving domains without forgetting the previously learned ones.

DIL is applied to solve various realistic problems of computer vision such as the detection of objects from visual road scenes of cities in Europe, India, and the United States incrementally \cite{garg2022multi}; the detection of objects from visual road scenes of Germany in different weather conditions: clear, fog, rain and snow, incrementally \cite{mirza2022efficient}; and so on. A DIL is yet to be explored for audio tasks.

In this work, we aim to develop a universal ASC system that should perform well in all audio domains (locations) seen so far. We consider a realistic online or 'single-pass through the data',  where the model sees the data stream only once. The proposed approach is different from existing domain adaptation systems for ASC in different locations or from different devices \cite{drossos2019unsupervised, mezza2021unsupervised, tan2024acoustic, singh2021prototypical}, which transfers knowledge from a source domain to a target domain where only the performance of the target domain is considered. 
 
 In \cite{drossos2019unsupervised, mezza2021unsupervised}, the performance of the system is analyzed in both the source and target domains:  mismatched recording devices, after adaptation in an unsupervised manner. The approach in \cite{drossos2019unsupervised} requires suitably sized target domain data to train a Generative Adversarial Network (GAN) based adaptation model offline for approximately 300 epochs, while \cite{mezza2021unsupervised} compute the statistics of each frequency band of the whole testing data of the target domain to match with training data of the source domain.  However, an incremental learning setup implies a sequence of successive multiple tasks or domains rather than only source and target domains, to evaluate the performance of the method over a long period \cite{hou2019learning, mulimani2024class}.

In this paper, we directly update the statistics of the Batch Normalization (BN) layers of the model using only a few samples from the training data of the new domain, in sequence, via an adaptive momentum scheme  \cite{mirza2022norm}. 
This method was also used in DIL for weather conditions using images \cite{mirza2022efficient}. However, in the context of ASC, it was not yet investigated what kind of domain shift is raised from different locations and how much that affects performance of a model across multiple domains.  

The major contributions of this work are as follows:
\begin{itemize}
    \item We propose to update only statistics of the BN layers of a single model to learn acoustic scenes in incremental domains 
    using a few samples, without the need for large training data to learn acoustic scenes from new locations.
    \item We investigate the performance of the proposed DIL approach to classify acoustic scenes from different European cities sequentially.
    \item 
    We also investigate the adaptation ability of the DIL model with a significantly large domain shift, using acoustic scenes from Korea in one of the incremental tasks.
\end{itemize}
The rest of the paper is organized as follows: Section 2 presents the behavior of BN layers in domain shift, notations, baselines, and the proposed DIL method for ASC. Section 3 introduces the datasets, implementation details, and results. Finally, conclusions are given in Section 4. 

\section{Domain-Incremental Learning}

Before presenting our DIL approach for ASC, we first explore the behavior of the BN layer in domain shift conditions.

\subsection{Batch Normalization layer}
The Batch Normalization (BN) layer is an important component of deep neural networks as it helps to stabilize the layer's distributions in the training phase \cite{ioffe2015batch}. 
Specifically, BN normalizes the input activations of each layer to have zero mean and unit variance as:
\begin{equation}
    \hat{\bm{h}} = \bm{\gamma} \left(\frac{\bm{h}-\bm{\mu}}{\sqrt{\bm{\sigma}^2+\epsilon}}\right)+\bm{\beta},
     \label{eq1}
\end{equation}
where mean $\bm{\mu}$ and variance $\bm{\sigma}^2$ are computed over a given mini-batch. Further, normalized features are affine-transferred with trainable scale $\bm{\gamma}$ and shift $\bm{\beta}$ parameters ($\epsilon$ is a small constant used for numerical stability). For the normalization during inference, a running mean $\bm{\hat{\mu}}$ and variance $\bm{\hat{\sigma}}^2$ are estimated via exponential moving average (EMA) at each training mini-batch $j$ as:
\begin{equation}
\begin{aligned}
    \bm{\hat{\mu}_j} \leftarrow \alpha{\bm{\mu_j}} + (1 - \alpha)\bm{\hat{\mu}_{j-1}}, \\ \bm{\hat{\sigma}}^2_{j} \leftarrow \alpha\bm{\sigma}^2_{j} + (1 - \alpha)\bm{\hat{\sigma}}^2_{j-1},
\end{aligned}  
 \label{eq2}
\end{equation}
where $\alpha$ is the momentum hyperparameter. It controls the balance between current and previously accumulated mean and variance over mini-batches. For brevity,  we refer mean $\bm{\hat{\mu}}$ and variance $\bm{\hat{\sigma}}^2$ as statistics hereafter.

Specifically, the behavior of the BN layer is different during the training and inference phases. In the training phase, statistics of the BN layer are updated during data forwarded through the neural network, forward pass, in mini-batches using Eq. (\ref{eq2}). On the other hand, in the inference phase, statistics obtained during training are fixed and used in Eq. (\ref{eq1}) for normalization. These statistics either improve the performance of the BN layer when both training and testing data have similar distributions or degrade the performance of the BN layer when training and testing data have mismatched distributions or shifted domains.

\subsection{DIL setup and notations}

In our DIL setup, a sequence of ASC tasks is presented to the model and these tasks represent the datasets from different domains: $\mathcal{D}_1, \mathcal{D}_2, ..., \mathcal{D}_t$. The model learns each task, i.e., $\mathcal{D}_t$ in our case, at incremental time step $t$. A domain $\mathcal{D}_t$ is an acoustic scene dataset collected in a particular geographic location composed of audio clips and corresponding class labels. All domains share the same classes. We aim to train a single-model $\mathcal{M}$ that learns to classify the same acoustic scenes when domain or data distribution changes. During training, $\mathcal{M}$ follows a realistic setting where it sees a stream of samples only once, online, and quickly adapts to the new domain on the fly. More importantly, the performance of the $\mathcal{M}$ does not degrade on previous domains when it learns a new domain, unlike the domain adaptation case, in which the performance on the previous domain does not matter. Note that in this work we refer to $\mathcal{D}_t$ as task, domain, location, and dataset interchangeably.

\subsection{Baselines}

We construct a few standard baselines to compare with the proposed DIL: 
(1) \textit{Base}: a model is trained offline on the first (base) domain $\mathcal{D}_1$ and then it is frozen. The performance of the base model is evaluated in incremental domains without modifying any of its parameters, (2) \textit{Feature extraction (FE)}: the feature extractor component of the model is frozen after learning  $\mathcal{D}_1$. The classifier is updated in each incremental domain;  
(3) \textit{Fine-tuning (FT)}: a current model is fine-tuned on the new domain at each incremental time step. The model being trained incrementally;
(4) \textit{Disjoint}: a separate model is trained on each domain. For a fair comparison, the base model is trained offline and other models are trained either online or offline, depending on the experimental setup;
(5) \textit{Joint}: trains a model from all the data of the domains seen so far, breaking one of the constraints of the DIL.

\begin{figure}[!tbp]
  \centering
  \includegraphics[width=8cm, height=3.8cm]{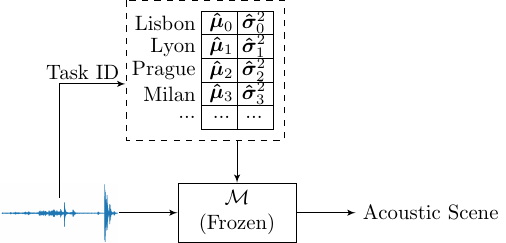}  
  \caption{Overview of the proposed online Domain-Incremental Learning approach. Inputs to the DIL model are the test sample and the task ID. The frozen model $\mathcal{M}$ uses domain-specific statistics to classify the acoustic scenes from a particular domain.}
     \label{fig:domain}
     \vspace{-10pt}
\end{figure}

\subsection{Domain-Incremental Learning approach}
An overview of our DIL approach is given in Fig.~\ref{fig:domain} At the initial time step $t=1$, the model $\mathcal{M}$ is trained on dataset $\mathcal{D}_1$. At each incremental time step, we only update the statistics of the BN layers of $\mathcal{M}$ in the forward pass without requiring back-propagation, and all other parameters of $\mathcal{M}$ are fixed.
The existing methods use fixed momentum $\alpha$ for the BN layer and its value is usually set to 0.1. In this work, we update $\alpha$ to adapt to the current domain as per the recommendation of \cite{mirza2022norm} with each input sample $k$ in an unsupervised manner as:
\begin{equation}
\begin{aligned}
\alpha_k = \alpha_{k-1}\times\omega,\\
    \alpha_k = \alpha_k + \delta,     
\end{aligned}  
 \label{eq3}
\end{equation}
where $\alpha_0=0.1$, $\omega\in\{0,1\}$ is a decay factor and constant $\delta$, $0<\delta<\alpha_0$, is a lower-bound of the momentum. $k=\{1, 2, ..., K\}$, $K$ is the total number of samples used to update the momentum, which is usually a small portion of the training data of the current domain. 
As $\alpha_k$ decays, the impact of the latter samples reduces and statistics effectively adapts to the samples of the current domain. We store these domain-specific statistics i.e., running mean and variance to the model. Specifically, we replace the $\alpha$ in Eq. (\ref{eq2}) with each value of $\alpha_k$ and check the performance of $\mathcal{M}$ to store the best statistics. 
Note that this adaptation is done in an unsupervised manner, which means there is no need for labeled data of the new domain--this means that it is possible to use a few samples either from the train or test dataset for the incremental learning of the new domain, hence solving the domain mismatch with a minimal amount of data.

During inference at each incremental phase, we apply domain-specific statistics to $\mathcal{M}$ to classify acoustic scenes of the current domain. The model expects input as a combination of task ID and test sample, similar to task-incremental learning \cite{li2017learning}. Task ID identifies the statistics of the corresponding task before classifying the test sample. We do not change any other parameters of the $\mathcal{M}$ except statistics of the BN layers. This enables us to recover the original performance of $\mathcal{M}$ in all steps by replacing the corresponding statistics. Therefore, $\mathcal{M}$ does not suffer from forgetting previous tasks when it learns a new task. Hereafter, we refer to the use of an adaptive momentum scheme for ASC in a sequence of domains as the online DIL (ODIL) approach.   
\section{Evaluation and Results}
\subsection{Datasets and training setup}
The domain $\mathcal{D}_1$ is composed of the TUT Urban Acoustic Scenes 2018 development dataset \cite{Mesaros2018_DCASE}. This dataset contains samples from 10 acoustic scenes in 6 different cities: Barcelona, Helsinki, London, Paris, Stockholm and Vienna. 
The remaining domains include the audio samples from 4 other cities of the TAU Urban Acoustic Scenes 2019 development dataset \cite{Mesaros2018_DCASE}. 
Specifically, $\mathcal{D}_2$, $\mathcal{D}_3$, $\mathcal{D}_4$ and $\mathcal{D}_5$ includes audio samples from Lisbon, Lyon, Prague, and Milan respectively.  
The domain $\mathcal{D}_6$ includes samples from the more diverse CochlScene dataset \cite{jeong2022cochlscene}, which contains samples from 13 acoustic scenes collected in Korea. However, particular recording location information is not available in this dataset. We consider for   $\mathcal{D}_6$ all samples from four acoustic scenes: bus, park, metro (named as subway), and metro station, which are also present in the other domains considered in this study.  

The total number of samples in the training and test splits of each dataset is given in  Table \ref{tab:SAMPLES}. Note that domains include audio samples from disjoint cities. We take a small number of $K$ samples randomly to update the momentum for the current domain. In this work, we select one sample per class, one shot, of the training data of the current European city. 

However, samples of Milan are only present in the test split of the TAU Urban Acoustic Scenes 2019 development dataset. For this specific case, we take the 10 samples from the test split itself to adapt ODIL to the $\mathcal{D}_5$. 
Because the other methods we use for comparison require going through the training samples, we only report the performance of Milan using the proposed ODIL method in subsection (\ref{res}). 

The domain $\mathcal{D}_6$ includes only 4 classes; for this domain we take two samples per class from the training data, for effective adaptation of ODIL to Korea. Therefore, in this case we use only 8 samples for ODIL, a much lower proportion of the data available for training the other methods. 

\begin{table}[]
  \caption{Number of training and testing samples in each domain. $K$  is the number of samples used to update the statistics of BN layers}
     \centering
   \begin{tabular}{l|c|c|c|c|c|c}
 \toprule
 Dataset splits&$\mathcal{D}_1$ & $\mathcal{D}_2$ & $\mathcal{D}_3$ & $\mathcal{D}_4$ & $\mathcal{D}_5$ & $\mathcal{D}_6$\\
 \midrule
 Train & 6122 & 1061 &976 & 1026&-&18674\\
Test & 2518 & 379 &464 & 414&410&2363\\
 K & - & 10 &10 & 10&10&8\\
 \bottomrule
 \end{tabular}     
     \label{tab:SAMPLES}
         \end{table}

\subsection{Implementation details and evaluation metrics}

We use the pre-trained PANNs CNN14 \cite{kong2020panns} architecture and fine-tune on the first domain $\mathcal{D}_1$ as the base system. Then, we use this fine-tuned CNN14 model in all other experiments. Input audio recordings are resampled to 32 kHz and
log mel spectrograms are computed using default settings provided in \cite{kong2020panns}. 

The model is trained using the SGD optimizer \cite{loshchilov2017sgdr} with a learning rate of 0.0001, a momentum of 0.9, and a mini-batch size of 32. The number of epochs to train the model on  $\mathcal{D}_1$ is set to 120. The baselines are trained at incremental time steps for 1 epoch for the online setting, and 120 epochs for the offline setting, respectively.
CosineAnnealingLR \cite{loshchilov2017sgdr} scheduler updates the optimizer in every epoch. The decay factor $\omega$ and constant $\delta$ are set to 0.94 and 0.05 respectively, based on the number of preliminary experiments.

We evaluate the performance of the model on all previously seen domains at each incremental step using average accuracy and forgetting. Average accuracy is the average of accuracies of the method over the current and all previously seen domains.  Average forgetting (Fr) is the average difference between the accuracy of the model for each domain at its learning iteration (first time the model learns this domain) and the accuracy of the model for the same domain at the current iteration (after learning the current domain). A lower Fr is better.

\subsection{Results}
\label{res}

\begin{figure*}[htbp!]
  \centering
  \subfloat[FE]{\includegraphics[width=5.5cm, height=3cm]{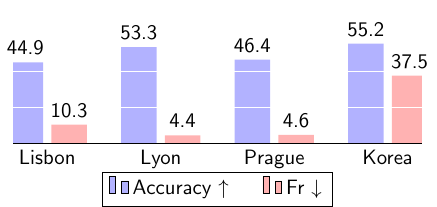}\label{fig:FE}}
  \hfill
  \subfloat[FT]{\includegraphics[width=5.5cm, height=3cm]{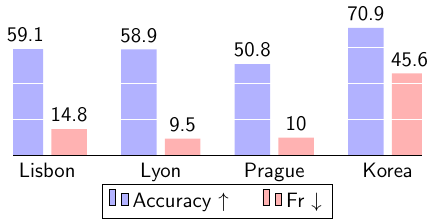}\label{fig:FT}}
  \hfill
  \subfloat[Disjoint]{\includegraphics[width=5.5cm, height=3cm]{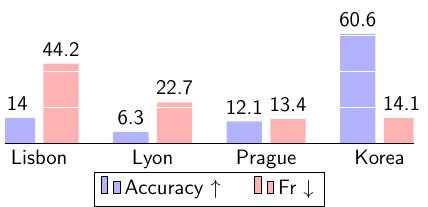}\label{fig:disj}}   
  \caption{Performance of the methods in online setting: accuracy at the current domain and average forgetting over previous domains.}
        \label{Fig:ONLINE}
     \vspace{-20pt}
\end{figure*}

\begin{figure*}[htbp!]
  \centering
  \subfloat[FE]{\includegraphics[width=5.5cm, height=3cm]{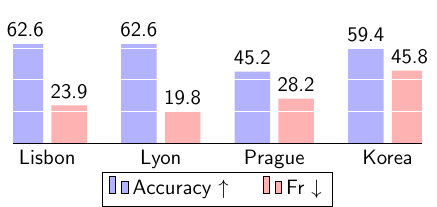}\label{fig:FEOFF}}
  \hfill
  \subfloat[FT]{\includegraphics[width=5.5cm, height=3cm]{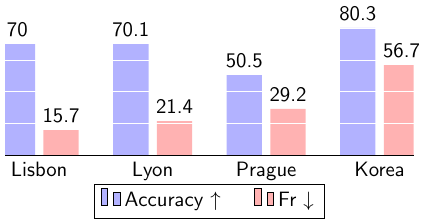}\label{fig:FTOFF}}
  \hfill
  \subfloat[Disjoint]{\includegraphics[width=5.5cm, height=3cm]{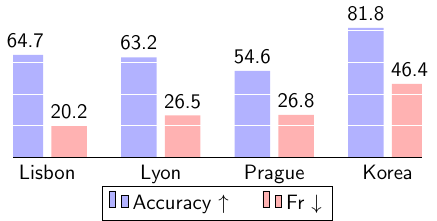}\label{fig:disjOFF}}   
  \caption{Performance of the methods in offline setting: accuracy at the current domain and average forgetting over previous domains.}
        \label{Fig:OFLINE}    
\end{figure*}

\begin{figure}[!tbp]
  \centering
  \includegraphics[width=5.5cm, height=3cm]{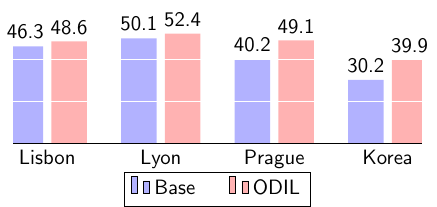}  
  \caption{Accuracy of the base and ODIL at different domains}
     \label{fig:ODIL}
     \vspace{-10pt}
\end{figure}

The model trained on the domain $\mathcal{D}_1$ achieved an accuracy of 59.8\% for $\mathcal{D}_1$. This model is used to learn the remaining domains using different methods: baselines and proposed ODIL approach for ASC. The average accuracy of each method over current and previously seen domains in online and offline settings is given in Tables \ref{tab:ONLINE} and  \ref{tab:OFLINE}, with  Fig.~\ref{Fig:ONLINE} and ~\ref{Fig:OFLINE} showing the accuracy of baselines on the current domain and the average forgetting over previously seen domains. 

In the online setting, the baseline systems are optimized for a single epoch on the training data of the current domain, hence these systems forget less on previous domains as compared to being trained in offline setting (120 epochs). On the other hand, in offline setting, baseline systems perform better on the current domain but exhibit significant forgetting on previous domains. This is visible in  Fig.~\ref{Fig:OFLINE} where forgetting consistently increases in incremental domains.

\begin{table}[]
 \caption{Average accuracy of the online methods over current and all previously seen domains.}
     \centering
   \begin{tabular}{l|c|c|c|c|c|c}
 \toprule
 Method & \makecell{$\mathcal{D}_1$\\6 cities} & \makecell{$\mathcal{D}_2$\\Lisbon} & \makecell{$\mathcal{D}_3$\\Lyon} & \makecell{$\mathcal{D}_4$\\Prague} & \makecell{$\mathcal{D}_5$\\Milan} & \makecell{$\mathcal{D}_6$\\Korea}\\
 \midrule
 Base & 59.8 & 53.1 &52.1 &49.1 &46.3 & 43.7\\
 \midrule
 FE & 59.8 & 47.2 &49.8 &47.8 & - & 21.9\\
 FT & 59.8 & 52.1 &52.9 &49.6 & - & 23.4\\
Disjoint & 59.8 & 14.8 &11.6 & 13.0&-&19.3\\
 \midrule
ODIL & 59.8 & \textbf{54.2} &\textbf{53.6} & \textbf{52.5} & \textbf{50.5} & \textbf{48.8}\\
 \midrule
 Joint & {59.8} & {56.0} &{58.8} & 55.9 &-&48.2\\
 \bottomrule
 \end{tabular}     
     \label{tab:ONLINE}
        \end{table}
\begin{table}[]
 \caption{Average accuracy of the offline methods over current and all previously seen domains.}
     \centering
   \begin{tabular}{l|c|c|c|c|c|c}
 \toprule
 Method & \makecell{$\mathcal{D}_1$\\6 cities} & \makecell{$\mathcal{D}_2$\\Lisbon} & \makecell{$\mathcal{D}_3$\\Lyon} & \makecell{$\mathcal{D}_4$\\Prague} & \makecell{$\mathcal{D}_5$\\Milan} & \makecell{$\mathcal{D}_6$\\Korea}\\
 \midrule
 Base & 59.8 & 53.1 &52.1 &49.1 &46.3&43.7\\
 \midrule
 FE & 59.8 & 49.3 & 48.5&36.4& -&21.2\\
 FT & 59.8 & 57.1 &52.3 &40.8 & - & 25.6\\
Disjoint & 59.8 & 52.2 & 44.9& 40.6&-&27.7\\
 \midrule
ODIL & 59.8 & \textbf{54.2} &\textbf{53.6} & \textbf{52.5} & \textbf{50.5} & \textbf{48.8}\\
 \midrule
 Joint & {59.8} & {58.5} &{65.7} & 60.5 &- &58.5\\
 \bottomrule
 \end{tabular}     
     \label{tab:OFLINE}
         \end{table}
     
The base model trained on the domain $\mathcal{D}_1$ is used to classify the acoustic scenes in incremental domains without changing any of its parameters, therefore it achieves zero-forgetting on previous domains. However, it does not put any effort into adapting to the current and upcoming domains and can fail to classify acoustic scenes of highly mismatched domains. 
Fig. ~\ref{fig:ODIL} shows the accuracy of the base and proposed ODIL for ASC in each current domain. It can be seen that the base model achieves poor performance on $\mathcal{D}_4$: Prague and $\mathcal{D}_6$: Korea. In comparison, the 43.7\% from Table \ref{tab:ONLINE} reflects the average over all domains, including its base domain $\mathcal{D}_1$ on which its performance is high.

FE only updates the classifier in each incremental domain, therefore achieves less forgetting of previous domains compared to FT, in which all the layers of the model are updated.  However, it does not effectively adapt to the current domain.  FE achieves an accuracy of 55.2\% and 59.4\% in online and offline settings respectively,  for classifying the acoustic scenes in Korea at the cost of significant forgetting of all the previously learned European cities. This can be seen in Fig.~\ref{fig:FE} and Fig.~\ref{fig:FEOFF} as a very high Fr for Korea ($\mathcal{D}_6$), and the reduced average accuracy for FE in Tables \ref{tab:ONLINE} and  \ref{tab:OFLINE}. 
On the other hand, FT achieves better performance in classifying the acoustic scenes of the current domain but again at the cost of significant forgetting of previous domains, with a high Fr for Korea, as seen in Fig.~\ref{fig:FT} and ~\ref{fig:FTOFF}. The good performance of FT in Tables \ref{tab:ONLINE} and \ref{tab:OFLINE} in each domain is highly dominated by the accuracy of the current domain. 

In the case of a disjoint approach, a separate pre-trained CNN14 is fine-tuned on each domain in online and offline settings. In an online setting (one epoch), the model fails to achieve good performance, as seen in Fig.~\ref{fig:disj} and  Table \ref{tab:ONLINE}. In an offline setting it performs better, comparable to FT for Prague
and Korea. FT achieves better performance than disjoint if the feature distribution of the current domain matches with the $\mathcal{D}_1$, which seems to be the case for Lisbon and Lyon.

Based on the results for all methods, we can say that the distribution of the features from $\mathcal{D}_6$: Korea is significantly different from that of the previous domains, the European cities. This high distribution shift makes the baselines: FE, FT, and disjoint, biased toward Korea, struggling to generalize, and significantly forget European cities, because the new distribution overwrites all previous knowledge of the European cities. The number of training samples available for Korea is much higher than in any other incremental domain, which is one of the possible reasons for baselines to learn this current domain better and almost forget the previous domains.  It is also clear from Fig.~\ref{fig:ODIL} that the base approach achieves poor performance on Prague because the distribution of training data of $\mathcal{D}_1$ is probably different from the test samples of Prague. It is also evident in Fig.~\ref{fig:FE} and ~\ref{fig:FT}, that the performance of FE, and FT is not high for Prague compared to Lisbon and Lyon. However, learning of acoustic scenes in Prague does not affect the previously learned European cities as much as learning acoustic scenes in Korea.

The proposed ODIL approach for ASC is highly effective even if the distribution of the current domain changes significantly from the base domain $\mathcal{D}_1$. ODIL improves the performance for Prague by 8.9\%p (percentage point) and Korea by 9.7\%p as compared to the base. ODIL quickly adapts the statistics of BN layers by just seeing a few (unlabeled) samples, and without forgetting any of the previous domains. It outperforms all other methods by a large margin after learning to classify acoustic scenes from 6 incremental domains, with an average accuracy of 48.8\%.  The performance of ODIL is highlighted in bold in Tables \ref{tab:ONLINE} and  \ref{tab:OFLINE} and its performance is close to the joint model in all domains.

\section{Conclusion}
In this paper, we presented a simple but effective approach: online domain-incremental learning  for the classification of acoustic scenes from different locations.  ODIL increases the performance of the model effectively in highly mismatched domains without forgetting previously learned knowledge. ODIL does not require any excessive training or labeled data, and is therefore suitable in realistic deployments. Future work includes the investigation of more domain-specific features to use for DIL in addition to BN statistics.
\bibliographystyle{IEEEtran}
\bibliography{references}
\end{document}